\documentclass{article}

\usepackage{lineno,hyperref}
\usepackage{amsmath,amssymb}
\usepackage{graphicx}

\usepackage[utf8]{inputenc}
\usepackage[english]{babel}
\usepackage[T2A]{fontenc}

\usepackage{comment} 

\modulolinenumbers[10]
\modulolinenumbers[1]
\newcommand\nnfootnote[1]{%
  \begin{NoHyper}
  \renewcommand\thefootnote{}\footnote{#1}%
  \addtocounter{footnote}{-1}%
  \end{NoHyper}
}

\begin{document}
\title{Role of solitary states in forming spatiotemporal patterns in a 2D lattice of van der Pol oscillators}
\author{I.A. Shepelev \footnotemark[1], A.V. Bukh \footnotemark[1], S.S. Muni \footnotemark[2], V.S. Anishchenko\footnotemark[1]}

{\renewcommand\thefootnote{\fnsymbol{footnote}}%
  \footnotetext[1]{Department of Physics, Saratov State University, 83 Astrakhanskaya Street, Saratov, 410012, Russia}\footnotetext[2]{School of Fundamental Sciences, Massey University, Palmerston North, New Zealand}}
\maketitle
\begin{abstract}
The behaviour of the ensemble of coupled van der Pol oscillators is abundant when coupling parameters are changed. Incoherent and chimera-like regimes are observed at small values of the coupling strength parameter. Synchronization takes place with increasing of the coupling strength and coupling range parameters. Various chimera and solitary states are realised when the coupling strength parameter is sufficiently large. The single van der Pol oscillator is taken in the regime of relaxation oscillations. Therefore, the lattice may realise spiral wave and spiral wave chimera regimes. Besides, target wave and solitary state regimes are observed in the ensemble. Finally, the solitary state and solitary state chimera regimes are firstly shown for the lattice of coupled van der Pol oscillators.
\end{abstract}

\nnfootnote{Keywords: Spatiotemporal pattern formation, chimera state, van der Pol oscillator, spiral wave, spiral wave chimera, target waves, target wave chimera, solitary state, solitary state chimera, nonlocal coupling}
\nnfootnote{E-mail addresses: I.A. Shepelev (\url{igor\_sar@li.ru}), A.V. Bukh (\url{buh.andrey@yandex.ru}), S.S. Muni(\url{s.muni@massey.ac.nz}), V.S. Anishchenko (\url{wadim@info.sgu.ru})}
\date{}

\section*{\label{sec:intro}Introduction}

Exploring complex spatiotemporal structures in active media and networks of oscillator is one of the most actual directions in nonlinear dynamics. Usually, these structures are typical for nonequilibrium nonlinear active media, networks or ensembles of nonlinear oscillators. These systems can be described by a system of nonlinear ordinary differential equations \cite{kopell1973plane,winfree1974rotating,hagan1982spiral, keener1986spiral,vasiliev1987autowave,    tyson1987spiral, biktashev1989diffusion, barkley1990spiralwave, ermakova2005blood}. 
Appearance of the complex spatiotemporal structures in these system is usually accompanied by the complex oscillating or auto-wave dynamics. The intensive investigation of such regimes has started in the middle of the last century and continues to this day  \cite{kuramoto2003rotating, zhang2004spiral, ermakova2005blood, kuramoto2006meanfield, martens2010solvable, kuzmin2012deformation, gu2013spiral, tang2014novel, panaggio2015Bchimera, xie2015twisted, li2016spiral, weiss2017weakly, totz2018spiral}. Individual elements in the systems listed above have been diffusively coupled which each other, i.e., each element symmetrically interacts only with its closest neighbors.  "Second wind"  to the study of complex spatiotemporal structures has been got due to the discovery of so-called chimera states in 2002 in an ensemble of the Kuramoto phase oscillators with the nonlocal interaction between elements \cite{Kuramoto-2002,Abrams-2004}. The coupling nonlocality means  that each individual oscillator is symmetrically coupled with all adjacent oscillators from the neighborhood with a certain radius \cite{tanaka2003complex}. Most of the discovered chimera states have been found in ensembles with nonlocally coupled oscillators \cite{Omelchenko-2011, Zakharova-2014, Panaggio-2015, Maistrenko-2015, Bogomolov-2017,shepelev2018chimera}. However, there are a number of works where chimera states has been discovered in the system with the global coupling \cite{yeldesbay2014chimeralike, schmidt2014coexistence} and even local one \cite{laing2015chimeras, shepelev2018local}. Chimera states attract a lot of attention due to this fact that these states are realized in a large number of networks consisting of oscillator of different types. Perhaps they play an important role in living and technical systems consisting of a great number of coupled active elements. The recent works shows that chimeras can be observed in the real experiments  \cite{hagerstrom2012experimental,tinsley2012chimera,viktorov2014coherence} and are even realized in the brain dynamics \cite{bansal2019cognitive}.

Most of chimera states has been found in the 1D oscillatory ensembles \cite{Kuramoto-2002,Abrams-2004,laing2009dynamics,Omelchenko-2011,Zakharova-2014,sawicki2017chimera}.However they has been also discovered in 2D and even 3D networks \cite{martens2010solvable,hagerstrom2012experimental,omelchenko2012transition,Maistrenko-2015,shepelev2018double,shepelev2019variety}. The transition from an one-dimensional chain to a two-dimensional lattice can lead to appearance new chimera types. The best-known kind of chimera in a 2D lattice of oscillators is a spiral wave chimera (SWC), which has been discovered in the works \cite{kuramoto2003rotating,shima2004rotating}.  Moreover, the nonlocal interaction does not necessary for the SWC existence. They has been found even for the local coupling \cite{tang2014novel,li2016spiral,kundu2018diffusion}. Another type of chimera states in 2D lattices has been recently discovered in a 2D lattice of the van der Pol oscillators in the work \cite{bukh2019spiral}. It has been called a target wave chimera. This presents the target wave with an incoherence cluster within the wave center. Moreover, it has been shown that this cluster consists of a large number of so-called solitary states. The solitary state is characterized by a special behaviour of point elements that differs from the behavior of the other elements of the system. This regime  has ben observed in different ensembles with different topology \cite{maistrenko2014solitary,jaros2018solitary,shepelev2017solitary} and is typical for similar systems. It has been shown in \cite{semenova2018mechanism} that multistability induced by the coupling nonlocality can be a reason of the solitary states appearance in ensembles. Furthermore, it is shown in \cite{rybalova2018mechanism} that solitary states can group within a certain spatial region and form an incoherence cluster. This states has been called the solitary state chimera.

In the present paper  we study numerically the dynamics of a 2D lattice of  van der Pol oscillators with nonlocal interaction. This system has been investigated for the fixed parameters of coupling in \cite{bukh2019bspiral}. The typical regimes for this system are spiral waves, spiral wave chimeras and also target wave chimeras, which are observed for different values of the control parameters of individual oscillators. Now we study variety of spatiotemporal regimes, when the coupling parameters are varied and the control parameters are fixed. It has been revealed that solitary states are typical for the system under study. Moreover, they are of great importance in the formation of dynamic regimes when the nonlocality degree is high. We analyse in detail the features of the basic spatiotemporal patterns and transition between them when coupling parameters are varied.
Special attention is paid to the evolution of spatiotemporal structures, the nature of which is associated with the solitary states.

\section{\label{sec:system}System under study}

In our research the van der Pol (vdP) oscillator is chosen as an individual element of the model under study and is defined by the following system of ordinary differential equations (ODEs):
\begin{equation}
\begin{array}{l}
\dfrac{dx}{dt}  = y,\\
\dfrac{dy}{dt} = \varepsilon(1-x^2)y - \omega^2 x,
\end{array}
\label{eq:vdP_single}
\end{equation}
where $x $ and $y $ are dynamical variables. The parameter $\varepsilon$ determines the nonlinearity level, while the parameter $\omega$ is responsible for the oscillator frequency.  The value of $\varepsilon=0$ corresponds to the supercritical Andronov-Hopf bifurcation which is resulted in the limit cycle birth for $\varepsilon>0$. 

We consider the model of a spatially organized  ensemble of oscillators which presents a 2D regular  $N \times N $  lattice with an edge $N=100 $ and consists of nonlocally coupled vdP oscillators \eqref{eq:vdP_single}. This lattice is described by the following system of ODEs:
%
\begin{equation}
\begin{array}{l}
\dfrac {dx_{i,j}}{dt} = y_{i,j}  + \dfrac {\sigma} Q
\sum\limits_{\tiny \begin{aligned}& k=i-P \\[-4pt] & p=j-P\end{aligned}}^{\tiny \begin{aligned}& i+P \\[-4pt] & j+P\end{aligned}}
\left(x_{k,p} - x_{i,j}\right),
\\
\dfrac{dy_{i,j}}{dt} = \varepsilon (1- x_{i,j}^2) y_{i,j} - \omega^2x_{i,j} ,\\[8pt]
i,j=1,...N,
\end{array}
\label{eq:grid}
\end{equation}
%
The double index of dynamic variables $ x_{i, j} $ and $ y_{i, j} $ with $ i,j = 1, ..., N $ determines the position of an element in the two-dimensional lattice. All the oscillators are identical in parameters and each of them is coupled with all the lattice elements from a square with side $ (1 + 2P) $ in the center of which this element is located. The integer $ P $ defines the nonlocality of coupling and is called the interaction interval. The case of $P=1 $ corresponds to the local coupling, while $P=N/2 $ is the case of  global coupling, when each element interacts with the whole system. It determines a number of the neighbors $ Q = (1 + 2P) ^ 2-1 $, which each element is coupled with.  We also use the notion of the coupling range $r=P/N $ in analogy with classical works on chimeras. The coupling is introduced only to the first dynamical equation for $\dot x_{i,j} $.

In the present work we consider only  zero flux boundary conditions (Neiman type) which are defined as follows: 
\begin{equation}
\begin{split}
\begin{cases}
x_{0,j}(t) = x_{1,j}(t),\\
x_{i,0}(t) = x_{i,1}(t),\\
x_{N,j}(t) = x_{N+1,j}(t),\\
x_{i,N}(t) = x_{i,N+1}(t),
\end{cases}
\end{split}
~
\begin{split}
\begin{cases}
y_{0,j}(t) = y_{1,j}(t),\\
y_{i,0}(t) = y_{i,1}(t),\\
y_{N,j}(t) = y_{N+1,j}(t),\\
y_{i,N}(t) = y_{i,N+1}(t).
\end{cases}
\end{split}
\label{eq:boundaries}
\end{equation}

Initial conditions for all considered cases are a multitude of random values of the variables with a uniform distribution within $x_0 \in [-1;1],~y_0 \in [-1;1] $. The system equations are integrated using the Runge-Kutta 4th order method with the step $dt=0.001$. All the regimes under study are obtained after the transient process of $t_{\rm trans}=20000 $ time units.

\section{Basic dynamical regimes of the lattice (\ref{eq:grid})}

\begin{figure}[!ht]
\centering
  \includegraphics[width=1.\linewidth]{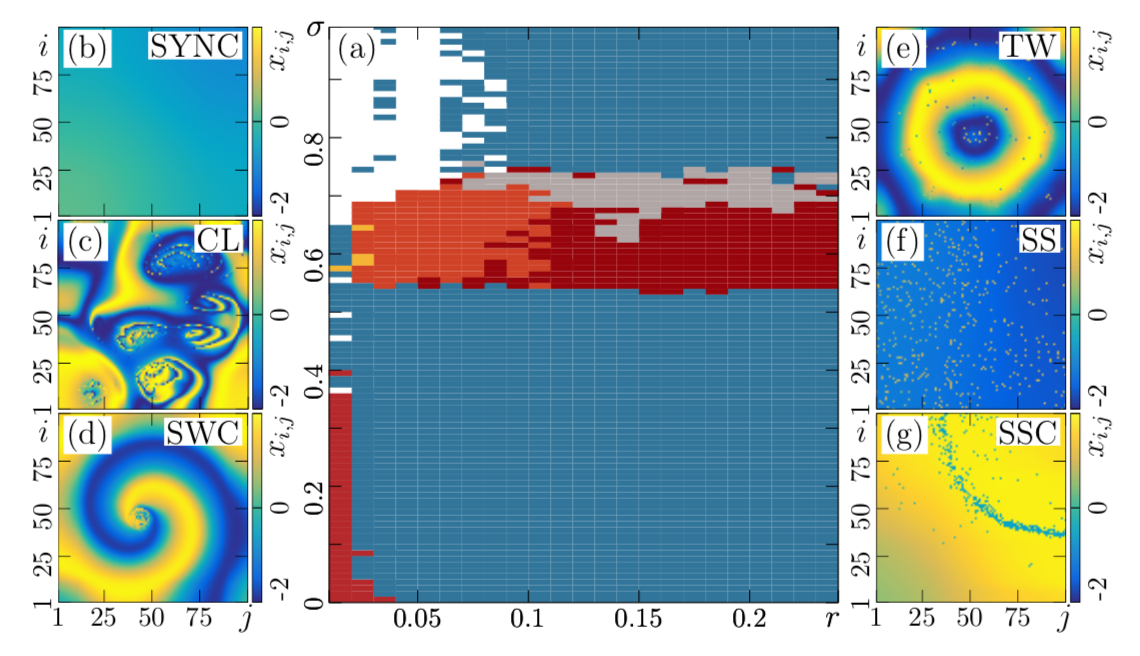}
\caption{(Color online) Diagram of the regimes for the lattice \eqref{eq:grid} in the ($r,~\sigma$) parameter plane for $\varepsilon=2.1 $ and $\omega=2.5$.
Region INC corresponds to complete incoherence; region SYNC relates to complete or partial synchronization; CL is the region of existence of chimera-like structures; regions SW and SWC correspond to spiral waves and spiral wave chimeras accordingly; TW is the region of target wave chimeras; region SS relates to the regime of solitary state; and solitary state chimeras are observed in region SSC. The regions of coexistence of different regimes are shown by alternating strips of the corresponding colors (tones). The regimes are illustrated by corresponding snapshots of the system states at $r=,\,\sigma=$ (b), $r=0.05,\,\sigma=0.01$ (c), $r=,\,\sigma=$ (d), $r=,\,\sigma=$ (e), $r=,\,\sigma=$ (f), and $r=,\,\sigma=$ (g).}
\label{fig:regime_map}
\end{figure}

Now we start to explore the dynamics of the lattice \eqref{eq:grid} when the parameters $\sigma$ and $r $ are varied and the control parameters $\varepsilon=2.1 $ and $\omega=2.5 $ are fixed. 
We plot a regime diagram for the lattice \eqref{eq:grid} in the ($r,~\sigma$) parameter plane within the ranges $\sigma \in [0;1]$ and $r \in (0;0.27] $ as shown in Fig.~\ref{fig:regime_map}(a). 
A sequence of randomly distributed initial conditions within the intervals $x_{i,j} \in [-1;1]$ and $ y_{i,j} \in [-1;1]$ is used to construct the diagram. The dashed regions correspond to the case of multistability in the lattice, when two different steady regimes are observed for various sets of the initial conditions.
%

\begin{figure}[!ht]
\centering
\parbox[c]{.4\linewidth}{ 
  \includegraphics[width=\linewidth]{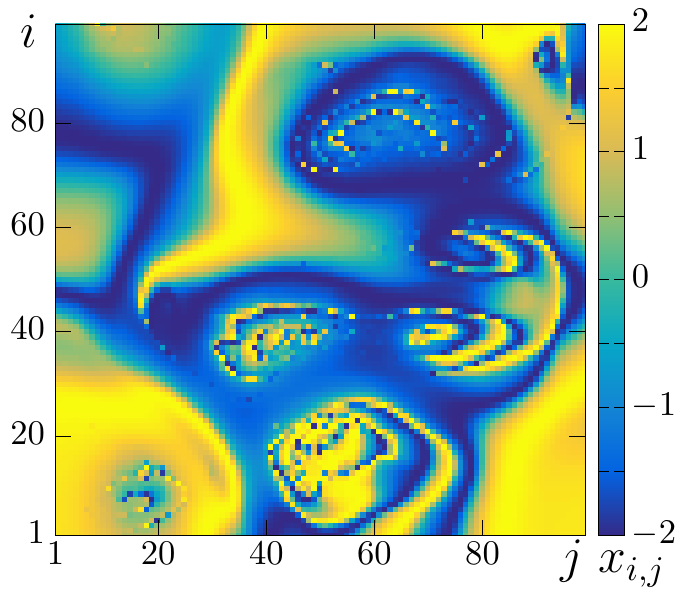}
    \vspace{-9.5mm} \center (a)
}
\parbox[c]{.4\linewidth}{
\includegraphics[width=\linewidth]{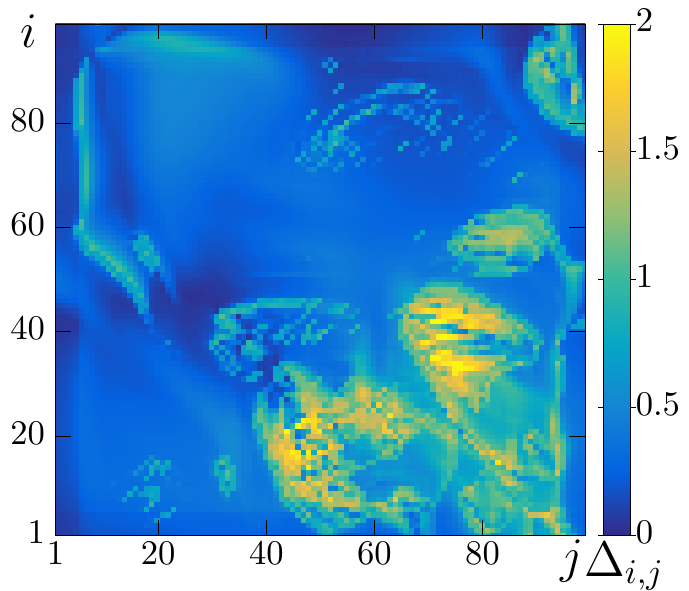}
\vspace{-9.5mm}\center (b)
}
\caption{(Color online) Chimera-like regime (CL) (region CL in  the regime diagram in fig.\ref{fig:regime_map}) in the lattice \eqref{eq:grid} for $r=0.05 $ and $\sigma=0.01 $. The snapshot of the system state (a), and the RMSD spatial distributions $\Delta_{i,j}$ (b). Parameters: $\varepsilon=2.1 $, $\omega=2.5$, $N=100 $} 
\label{fig:CL}
\end{figure}
%

The incoherent regime is realized when the coupling range $r $ is short and values of the coupling strength $\sigma $ are sufficiently small.  This regime is denoted by letters INC in the regime diagram. 
The chimera-like structures are realized in the system~\eqref{eq:grid}, when values of the coupling strength and the coupling range increase. This regime is illustrated in a snapshot of the system state in Fig.\ref{fig:CL} and denoted by letters CL in the regimes diagram in Figs.~\ref{fig:regime_map}(a),(c). The regime is characterized by the coexistence of coherent clusters with a slow changing of the amplitudes, and incoherent clusters with embedded coherent and incoherent clusters. These incoherent clusters represent the structures similar to fractals (see Fig.~\ref{fig:CL}(a)). 

The root-mean-square  deviation (RMSD) $\Delta_{i,j} $ is applied to quantify  the regime and is calculated as follows:
\begin{equation}
\begin{array}{l}
\Delta_{i,j}=\sqrt{\langle(x_{i,j}-x_{i+1,j+1})^2\rangle}
\end{array}
\label{eq:RMSD}
\end{equation}
The spatial distribution of the RMSD takes the maximal values in the incoherence clusters and minimal ones in the coherence clusters as shown in Fig.~\ref{fig:CL}(b).
The synchronous regime  takes largest region in the ($r,~\sigma$) parameter plane. This regime is realised with $r>0.01$ and $\sigma\in[0:0.6)\cup(0.75:1]$ (see Fig.~\ref{fig:regime_map}(b)). All the elements of the lattice oscillate periodically. The instantaneous states of oscillators are the same for the case of complete synchronization, while for the partial synchronization the instantaneous states are almost the same for oscillators located closely from each other and can be different for far located elements. This regime is denoted by letters SYNC in the regime diagram and represented in Fig.\ref{fig:regime_map}(b).

A new interesting regime appears, when the coupling strength is sufficiently large.  The target wave chimeras are realized when $r<0.12$ and $0.55<\sigma<0.7$ (TW in Fig.\ref{fig:regime_map}(a),(e)). This regime represents concentric waves with solitary states, which forms the incoherence cluster of chimera, but can also exist outside this cluster in a common case. Oscillators corresponding to the solitary states are characterized by a attractor which differs form the one of the oscillators from the coherence region.The spiral waves and spiral wave chimeras are observed, when the coupling range is  $r<0.12$ and the coupling strength  $\sigma>0.7$. However, the spiral waves can also be realized for lower values of $\sigma $ and the local coupling ($P=1 $). This structures have already been found and studied in detail in the system~\ref{eq:grid} in the work~\cite{bukh2019bspiral}. For this regimes there is a center of the wave rotation from where the wavefront begins its propagation. Regions of the spiral waves and spiral wave chimera states are denoted by letters SW and SWC correspondingly in the regime diagram in fig.\ref{fig:regime_map}(a). The wavelength of the target wave  chimera is elongated with increasing the coupling range $r$ within $0.55<\sigma<0.7$ and this regime evolves to the  solitary state regimes when the coupling range becomes $r>0.12$. An example of this structure is presented in Fig.\ref{fig:regime_map}(f) and corresponds to region SS in the regime diagram in Fig.\ref{fig:regime_map}(a). Increase of the coupling strength ($\sigma > 0.7 $) for the long coupling  ranges $r>0.9$ leads to gradual gathering solitary states to a certain spatial cluster, and to a transition to the solitary state chimera regime, which is denoted as SSC in Fig.\ref{fig:regime_map}(a),(g). We consider last regimes in detail below.

\subsection{Spiral wave and spiral wave chimera}

\begin{figure}[!ht]
\centering
\includegraphics[width=\linewidth]{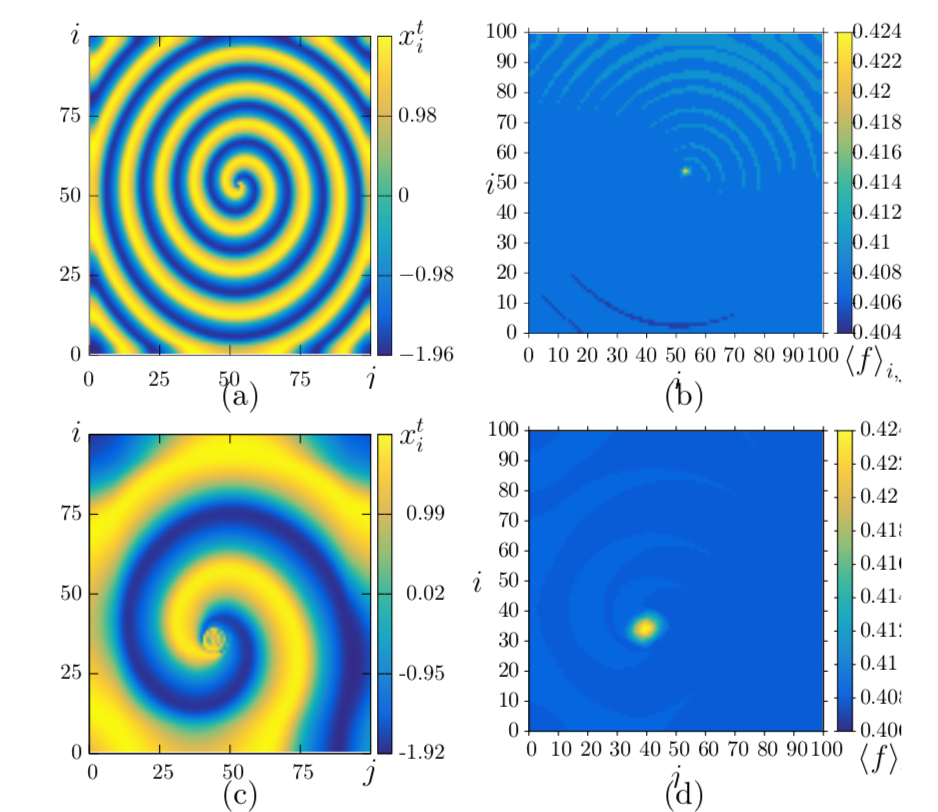}
\caption{(Color online) Spiral wave ((a), (b), region SW in the regime diagram in fig.\ref{fig:regime_map}) for $r=0.08,\sigma=0.8$; and spiral wave chimera ((c), (d), region SWC in the regime diagram in fig.\ref{fig:regime_map}). (a) and (c) present snapshots of the system state, (b) and (d) show spatial distributions of the mean frequencies $\langle f\rangle_{i,j} $. Parameters: $\varepsilon=2.1$, $\omega=2.5$, $N=100$}
\label{fig:SWs}
\end{figure}

Spiral wave regimes are widely known and has been found in various oscillatory lattices with different nature of the individual elements. The spiral wave shown in Fig~\ref{fig:SWs},(a) is obtained for the case when the coupling is close to local ($r=0.02$). Its spatial distribution of the mean frequency is shown in fig.{fig:SWs},(b). The element of the rotation center has the frequency slightly higher than the rest oscillators.

Increase of the coupling radius $r$ leads to forming the  incoherence core. The core arises right around the rotating center of the spiral wave. An example of this structure is pictured in~Fig~\ref{fig:SWs},(c). Note, that the spiral wave chimera is realized in the system \eqref{eq:grid} even for the randomly-distributed initial conditions. The oscillator behavior in the core of the spiral waves becomes chaotic, while the elements outside this core continue to oscillate periodically. The mean frequency of the oscillators within incoherent core increase in comparison with the frequency of oscillators in the coherent region. Moreover, peculiar ''bell-like'' spatial distribution of the mean frequency takes place. 
This fact is illustrated in the spatial distribution of the mean frequency in Fig~\ref{fig:SWs},(d). 

\subsection{Target wave chimera}

\begin{figure}[!ht]
\centering
\parbox[c]{.4\linewidth}{ 
  \includegraphics[width=\linewidth]{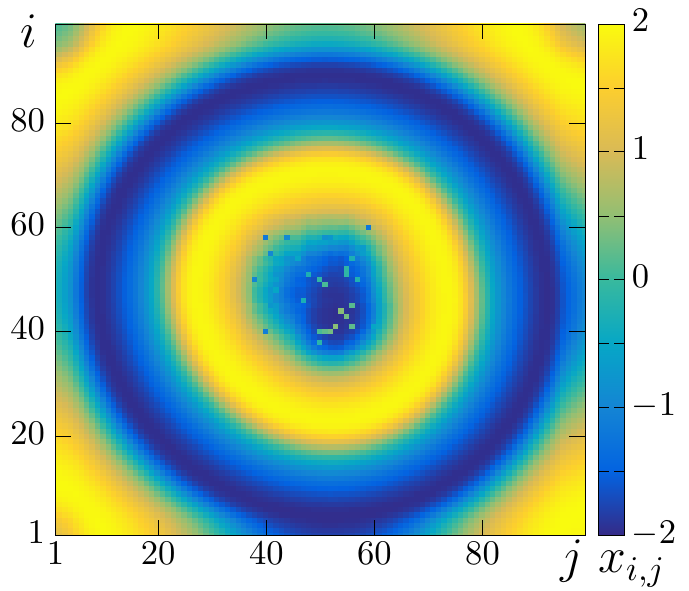}
    \vspace{-9.5mm} \center (a)
\includegraphics[width=\linewidth]{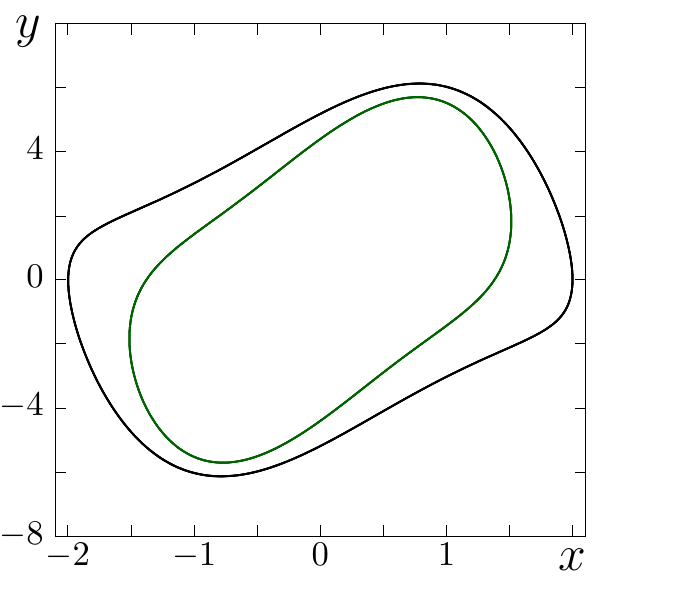}
\vspace{-9.5mm}\center (c)
\includegraphics[width=\linewidth]{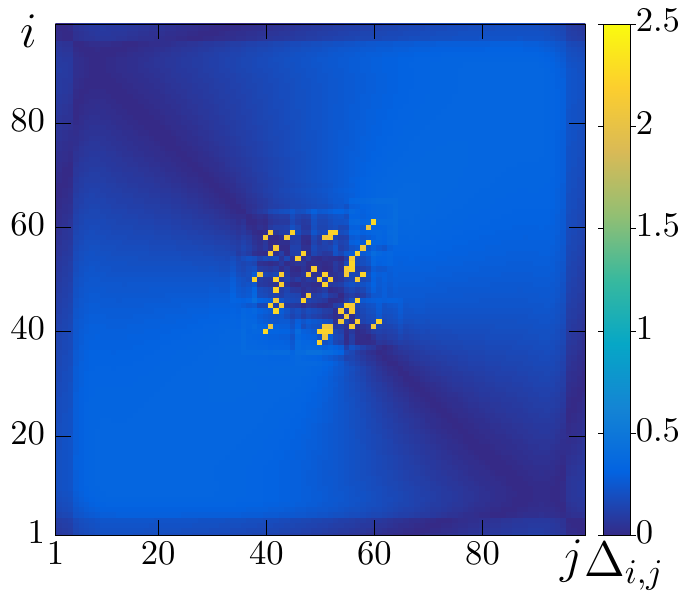}
\vspace{-9.5mm}\center (e)
}
\parbox[c]{.4\linewidth}{
  \includegraphics[width=\linewidth]{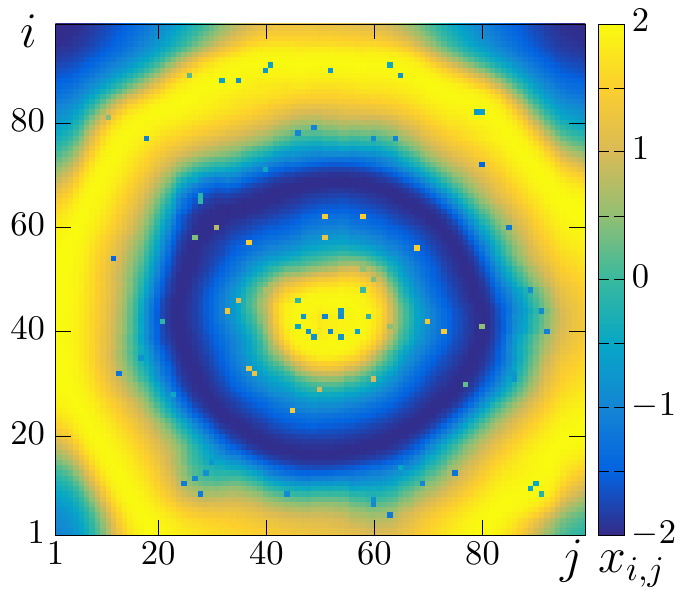}
   \vspace{-9.5mm} \center (b)
\includegraphics[width=\linewidth]{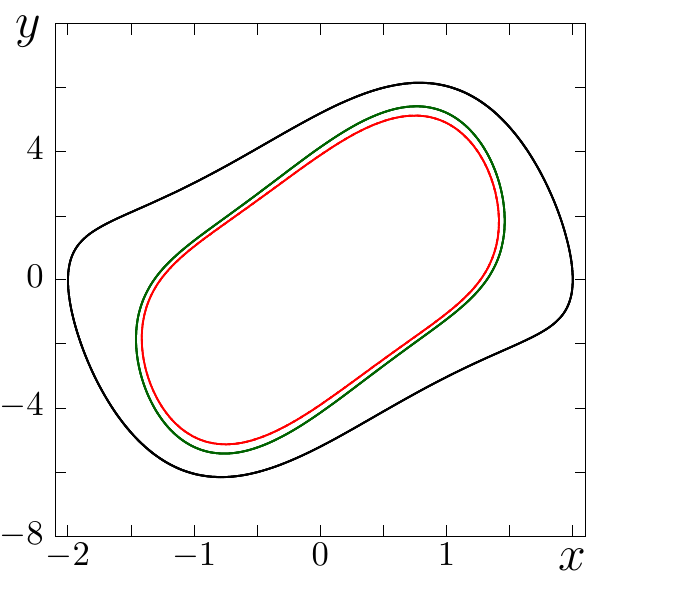}
 \vspace{-9.5mm} \center (d)
\includegraphics[width=\linewidth]{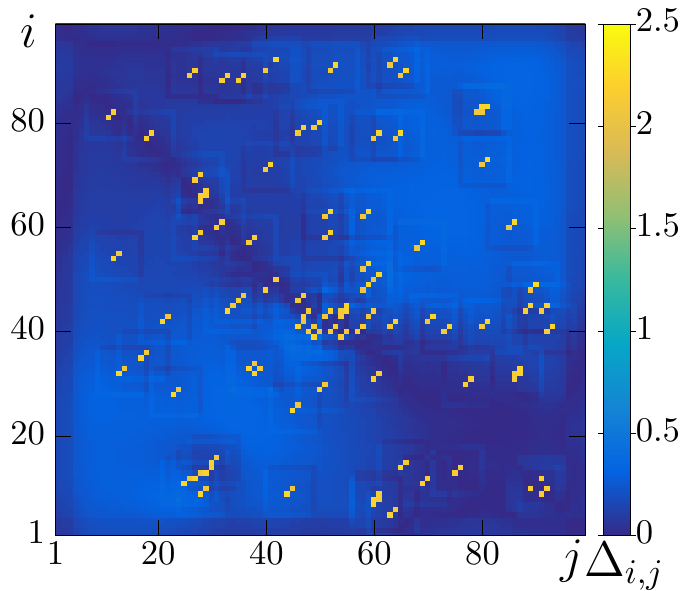}
 \vspace{-9.5mm} \center (f)
}
\caption{(Color online) Target wave chimera (TWC) (region $TW $ in  the regime diagram in fig.\ref{fig:regime_map}) in the lattice \eqref{eq:grid} for $r=0.04 $ and $\sigma=0.65 $.(a), (c), and (e) correspond TWC without the solitary state inside the coherence cluster, while (b), (d), and (f) are for TWC with the solitary state inside the coherence cluster. (a) and (b) are snapshots of the system state, (b) phase portrait projections for oscillators with indexes $i=1,~j=1 $ (coherence cluster, black line) and  $i=46,~j=47 $(incoherence cluster, green line), (c) phase portrait projections for oscillators with indexes $i=1,~j=1 $ (coherence cluster, black line), $i=48,~j=40 $ (incoherence cluster, green line), and $i=9,~j=28 $ (solitary state inside the coherence cluster, red dashed line), (e) and (f) show spatial distributions of the RMSD $\Delta_{i,j} $. Parameters: $\varepsilon=2 $, $\omega=2.5$, $N=100 $} 
\label{fig:TW}
\end{figure}

One of the basic dynamical regime in the two dimensional lattice \eqref{eq:grid} of coupled oscillators is target wave chimera (TW). It's existence is associated with the non-local coupling ($r0.01>1$), where $r$ denotes the coupling range. This regime is observed in the lattice \eqref{eq:grid} for $r>0.01$. A typical example of the TW is shown in fig.\ref{fig:TW}(a). It can be seen that the wavefront propagates from a certain wave center in the form of concentric circles in the direction of edges. At the same time instant, the core of the wave transforms into a spatial region with an incoherent spatial distribution forming the TW incoherence cluster. The remaining part of the lattice is characterized by a smooth spatial profile and forms the coherence cluster. It has been shown in \cite{bukh2019spiral} that the incoherence cluster presents a spatial region with a large number of the solitary states. 

The mechanism behind the appearance of solitary states as shown in \cite{rybalova2018mechanism} is associated with bistability or even multistability, which is a consequence of non-local coupling. It leads that elements in the solitary state regime are characterized by another attractor other than the synchronized elements. Fig.\ref{fig:TW}(c) presents the phase portrait projections of elements from the incoherence cluster (in the regime of solitary states) and  from the coherence cluster. It is seen that oscillations of these elements corresponds to different attractors. It should be noted that all the elements of the coherence cluster are characterized by the same attractor as well as the elements which are in the solitary state regime. However, the target wave chimera with incoherence cluster localized around the center is a particular case. Generally the state realized in region TW in the regime diagram in fig.\ref{fig:regime_map} is the target wave chimera with randomly spatially distributed solitary states in the coherence cluster. An example of this state is represented in fig.\ref{fig:TW}(b). For a lattice with a particular set of parameters, we see solitary states evolving over time.  In a snapshot of the system state they look like dots with a color different from the color of a coherence cluster region. 

We also note that these states noticeably impact the wavefront, which gets slightly distorted after passing through the solitary states. It is important that the position of the solitary state does not change over time. Fig.\ref{fig:TW}(d) demonstrates different phase portrait projections of the oscillators from the incoherence and coherence clusters, and of an element in the solitary state regime. 
One can see that coexistence of at least three attractors takes place for this case, i.e. a reason of the solitary state occurrence is also appearance of multistability. Moreover, the phase portrait projections for oscillators in the solitary state regime are very similar. 
It enables us to assume that the nature of the solitary state appearance is the same both inside the incoherence cluster and outside it in the lattice \eqref{eq:grid}. 

Spatial distributions of the RMSD for the both types of target wave chimeras are presented in Figs.\ref{fig:TW}(e)and (f). The coherence cluster with the auto-wave dynamics is characterized by minimal values of the RMSD $\Delta_{i,j} $, while the solitary states and incoherence cluster are characterized by the maximal values of $\Delta_{i,j} $. This is due to the fact that the adjacent elements in the auto-wave region oscillate regularly with  similar instantaneous states. At the same time, the elements in the solitary state regime  oscillate with different instantaneous phases and their instantaneous states may differ significantly. It should be noted, that the frequency of oscillation for all the elements is almost the same, and oscillations themselves demonstrate the regular periodic behavior, which is confirmed by a zero value of the maximal Lyapunov exponent $\lambda_{max}=0.00012 $.

\subsection{Solitary state regime}

\begin{figure}[!ht]
\centering
\parbox[c]{.4\linewidth}{ 
  \includegraphics[width=\linewidth]{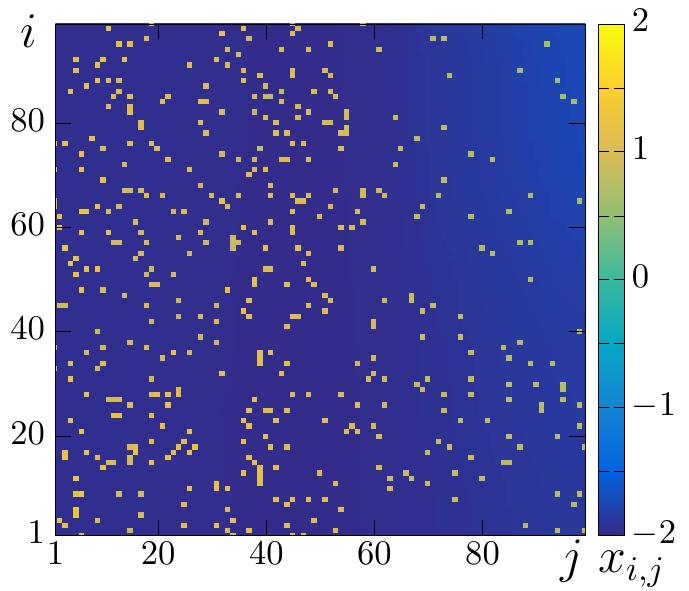}
    \vspace{-9.5mm} \center (a)
\includegraphics[width=\linewidth]{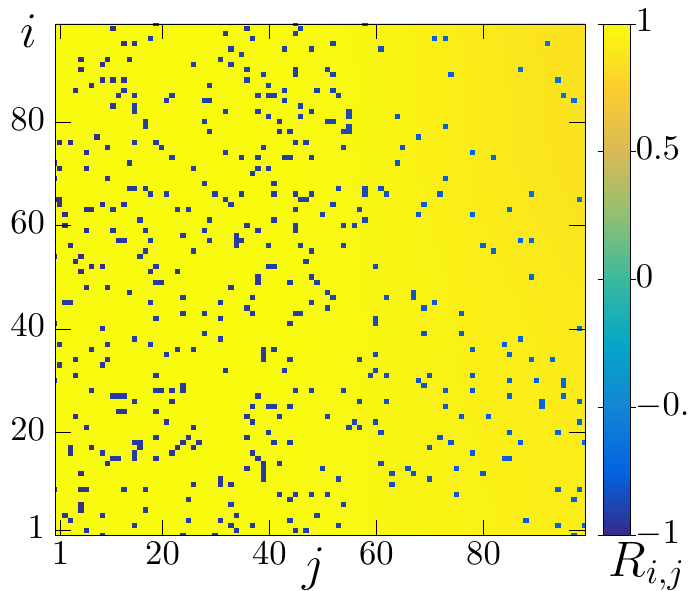}
\vspace{-9.5mm}\center (c)
}
\parbox[c]{.4\linewidth}{
  \includegraphics[width=\linewidth]{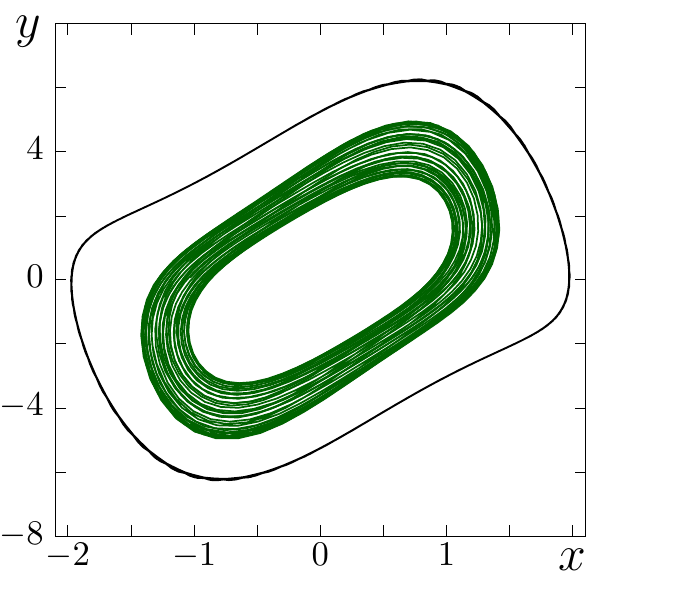}
   \vspace{-9.5mm} \center (b)
\includegraphics[width=\linewidth]{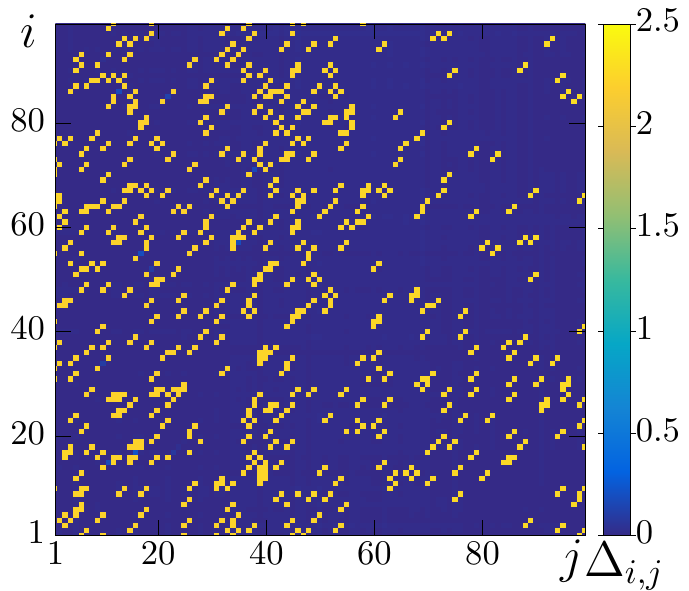}
 \vspace{-9.5mm} \center (d)
}
\caption{(Color online) Solitary state regime (SS) (region $SS $ in  the regime diagram in fig.\ref{fig:regime_map}) in the lattice \eqref{eq:grid} for $\sigma=0.57 $ and $r=0.3$.(a) is a snapshot of the system state, (b) phase portrait projections for oscillators with indexes $i=1,~j=1 $ (synchronous region, black line) and  $i=39,~j=28 $(solitary state, green line), (c) presents a spatial distribution of the crosscorrelation $R^{1,1}_{i,j} $ of the oscillator from the synchronous cluster with the other (d) shows spatial distributions of the RMSD $\Delta_{i,j} $. Parameters: $\varepsilon=2 $, $\omega=2.5$, $N=100 $}
\label{fig:SS}
\end{figure}

Now we consider evolution  of the target wave chimera when the coupling range increases. The increase in coupling strength leads to the transition of the system \eqref{eq:grid} from TWC to the regime of randomly spatially distributed solitary states (region $SS$ in the regime diagram in fig.\ref{fig:regime_map}). This regime is characterized by the almost synchronous behavior of most of the oscillators (instantaneous amplitude and phases of adjacent oscillators are very similar) while a part of oscillators demonstrate different dynamics from the synchronous part. However, complete syncronization is not observed, the instantaneous spatial profile is smooth but not flat.  Oscillators in the solitary state regime are randomly distributed within the lattice. An example of this state is shown in fig.\ref{fig:SS}(a).

The mechanism of the transition from region $TW$ to $SS$ is as follows. The wavelength of the target wave increases with growth of the coupling range $r $. For this reason the wave process disappears for large values of $r$. The shape of the spatial profile becomes smooth. At that instant, the solitary states which exists along with the target wave remain in the system. Their dynamics is changed too. The oscillations of an element in the solitary state regime are already quasi-periodic, and their phase portrait is of a torus, which is shown in fig.\ref{fig:SS}(b). The phase portrait projections looks similar to that of the chaotic attractor, but a zero value of the maximal Lyapunov exponent implies that oscillations behave regularly. A phase portrait projection of oscillators from the synchronous region corresponds to relaxation periodic oscillations and it noticeably differs from the solitary state.

We calculate the crosscorrelation $R^{k,l}_{i,j} $ of the selected $k,l $th oscillator (from the synchronous region) with the remaining elements ($1 \leq i,j \leq N$) for a quantitative analysis of the regime. It is calculated as follows:
\begin{equation}
\begin{aligned}
& R^{k,l}_{i,j} = \dfrac
{\langle \tilde x_{k,l} \tilde x_{i,j} \rangle}
{\sqrt{\langle \tilde x_{k,l}^2 \rangle
\langle \tilde x_{i,j}^2 \rangle}}, \\
& \tilde x_{i,j} = x_{i,j} - \langle x_{i,j} \rangle
\end{aligned}
\end{equation}
A spatial distribution of the crosscorrelation is represented in fig.\ref{fig:SS}(c). It shows that oscillators of the synchronous region are characterized by values of $R^{k,l}_{i,j} \approx 1$. It means that these elements oscillate in almost in-phase during all the observation time. On the other hand, elements in the solitary state regime have  values of the crosscorrelation $R^{k,l}_{i,j} \approx -1$. Thus, these elements oscillate in anti-phase to the synchronous oscillators, and almost in-phase with each other. The spatial distribution of the RMSD $\Delta_{i,j}$ is shown in fig.\ref{fig:SS}(d). The values of $\Delta_{i,j}$ for the synchronous region is close to zero, while oscillators in the solitary state regime have high values of the RMSD.

\begin{figure}[!ht]
\centering
\parbox[c]{.32\linewidth}{ 
  \includegraphics[width=\linewidth]{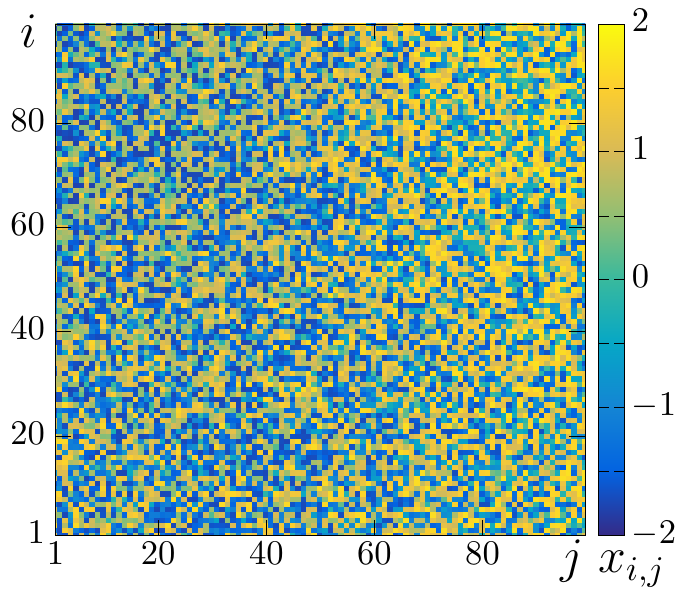}
    \vspace{-9.5mm} \center (a)
}
\parbox[c]{.32\linewidth}{
  \includegraphics[width=\linewidth]{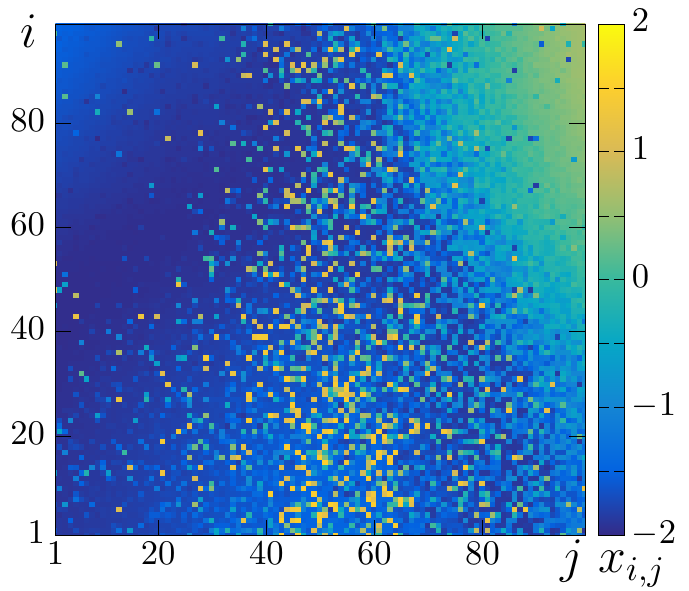}
   \vspace{-9.5mm} \center (b)
}
\parbox[c]{.32\linewidth}{
  \includegraphics[width=\linewidth]{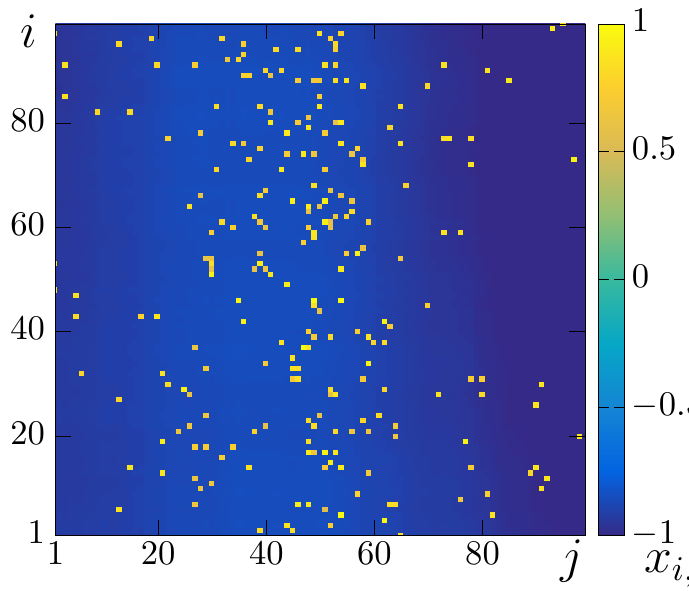}
   \vspace{-9.5mm} \center (c)
}
\caption{(Color online) Establishing the solitary state regime for the case of randomly distributed initial conditions. Snapshots of the system state at (a) $t=20 $ time units, (b) $t=40 $, and (c) $t=100 $
Parameters: $\sigma=0.57 $, $r=0.3$ $\varepsilon=2 $, $\omega=2.5$, $N=100 $} 
\label{fig:SS_evolution}
\end{figure}

Next, we show how the solitary state regime emerges from the randomly distributed initial conditions. The evolution in time of this regime is presented in fig.\ref{fig:SS_evolution}.
Adjacent elements begin to synchronize with each other after a short enough time (see fig.\ref{fig:SS_evolution}(a)). As it is seen in fig.\ref{fig:SS_evolution}(b), a large part of the oscillators are already synchronized. Solitary states becomes noticeable at this time moment. Fig.\ref{fig:SS_evolution}(c) corresponds to the steady solitary state regime, when most of the oscillators are synchronized with each other. Their oscillations correspond to the phase portrait shown in fig.\ref{fig:SS}(b). Other part of oscillators are in the solitary state regime with the other phase portrait. 

\subsection{Solitary state chimera}

\begin{figure}[!ht]
\centering
\parbox[c]{.32\linewidth}{ 
  \includegraphics[width=\linewidth]{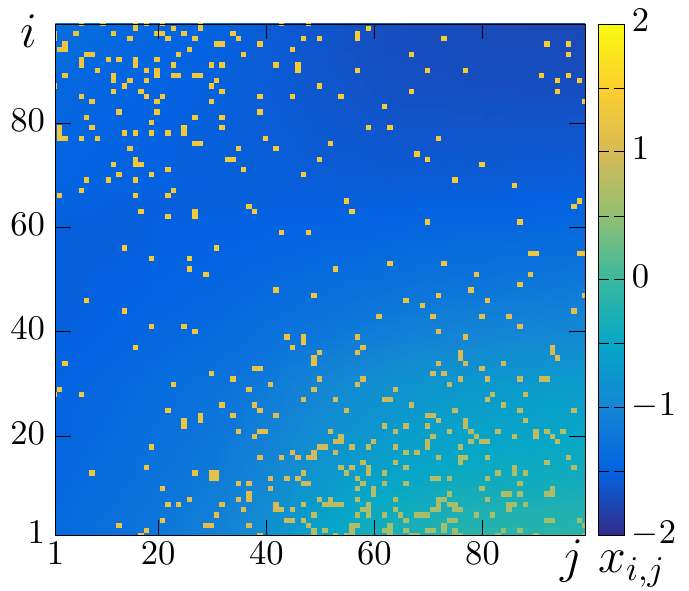}
    \vspace{-9.5mm} \center (a)
  \includegraphics[width=\linewidth]{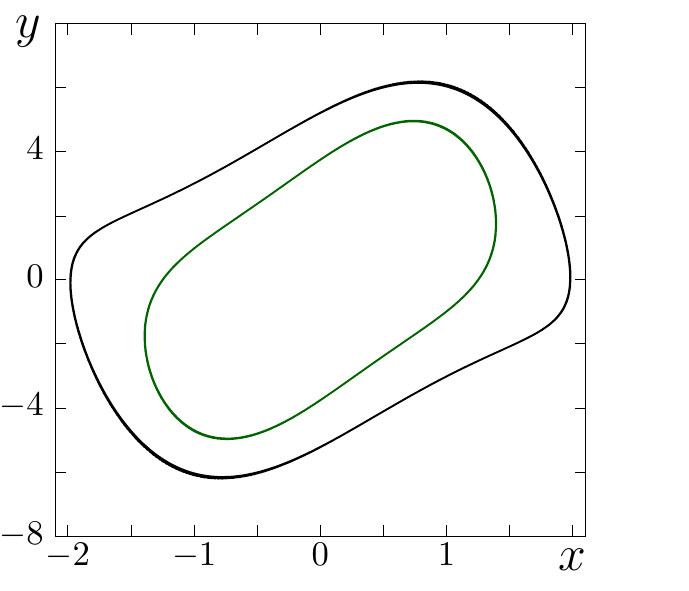}
    \vspace{-9.5mm} \center (d)    
}
\parbox[c]{.32\linewidth}{
  \includegraphics[width=\linewidth]{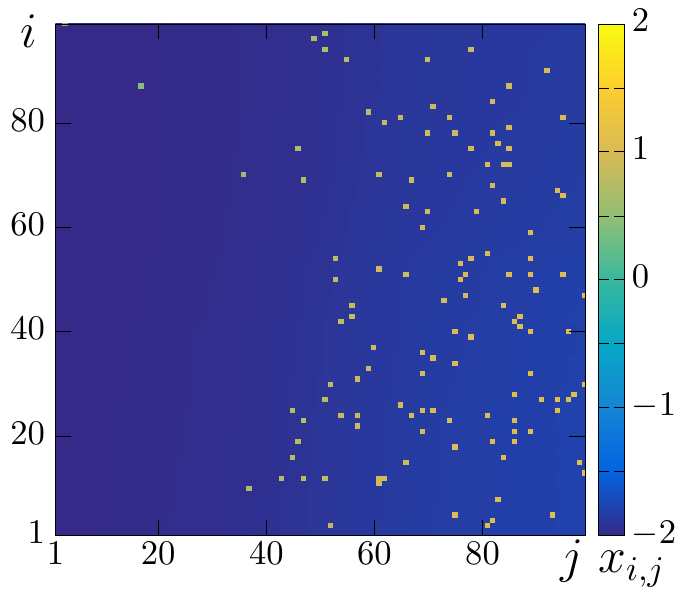}
   \vspace{-9.5mm} \center (b)
\includegraphics[width=\linewidth]{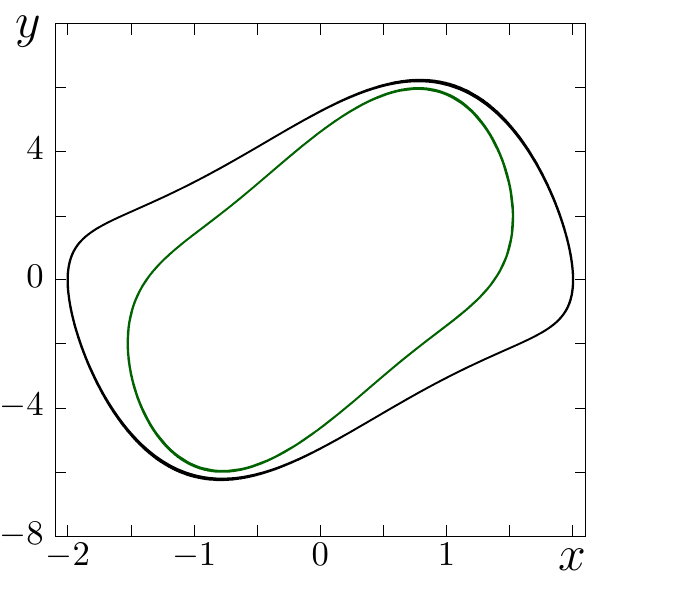}
    \vspace{-9.5mm} \center (e) 
}
\parbox[c]{.32\linewidth}{
  \includegraphics[width=\linewidth]{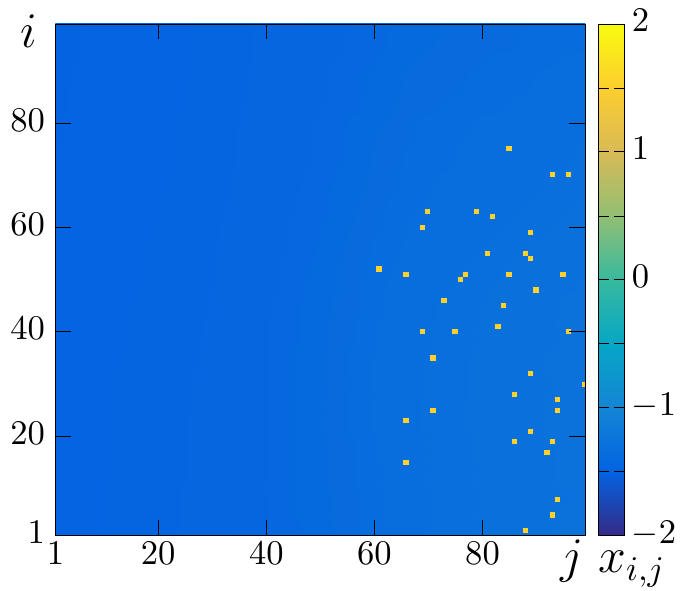}
   \vspace{-9.5mm} \center (c)
  \includegraphics[width=\linewidth]{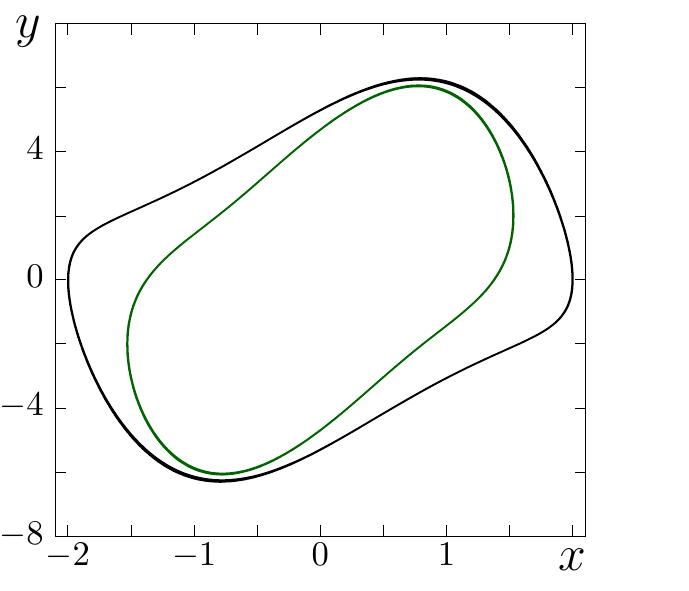}
    \vspace{-9.5mm} \center (f) 
}
\caption{(Color online) Transition from the solitary state regime ($SS $ region in the regime diagram in fig.\ref{fig:regime_map}) to the regime of solitary state chimera ($SSC $ region) for the same randomly distributed initial states. Snapshots of the system state  for (a)  $\sigma=0.612 $ , (b)  $\sigma=0.696 $, (c) $\sigma=0.726 $ and corresponding phase portrait projection for elements in the solitary state regime ($LC_{SS} $, green line) and the synchronous regime ($LC_{sync} $, black line) with indexes  (d) $i=40,j=78 $ and $i=40,j=77 $, (e) $i=2,j=85 $ and $i=1,j=85 $, (f) $i=51,j=85 $ and $i=52,j=85 $.
Parameters: $r=0.3 $, $\varepsilon=2 $, $\omega=2.5$, $N=100 $} 
\label{fig:SS-SSC}
\end{figure}

The solitary states begin to group within certain spatial region. Increasing the coupling strength leads to an interesting phenomenon. The solitary states  are distributed in space randomly and is shown  by $SS$ in the regime diagram in fig.\ref{fig:regime_map}. An example of this state is shown in fig.\ref{fig:SS-SSC}(a). Solitary states begin to approach a certain spatial region with increase in the coupling strength as shown in fig.\ref{fig:SS-SSC}(b). We also see that the region of the synchronous dynamics expands, while a number of the solitary states decreases. It is clearly seen in comparison between regimes in figs.\ref{fig:SS-SSC}(b) and \ref{fig:SS-SSC}(c). Thus, the regime realizied in the region $SSC$ of the regime diagram presents a certain type of the solitary state chimera. Further increase in the coupling strength leads to transition to the synchronous regime without solitary states.

We also consider evolution of the solitary state attractor with the increase in $\sigma $. Fig.\ref{fig:SS-SSC}(d) shows that when a value of the coupling strength $\sigma $ corresponds to $SS $ regime, the limit cycle of solitary state ($LC_{SS} $) is noticeably smaller than the limit cycle for oscillators of the synchronous region ($LC_{sync} $).  
Increase in the coupling strength leads to expansion of $LC_{SS} $ (see fig.\ref{fig:SS-SSC}(e)). This limit cycle almost approaches $LC_{sync}$ with increase in $\sigma$. Apparently, disappearance of solitary states is accompanied with tangency and merging of two limit cycles  ($LC_{SS} $ with $LC_{sync} $).

\begin{figure}[!ht]
\centering
\parbox[c]{.4\linewidth}{ 
  \includegraphics[width=\linewidth]{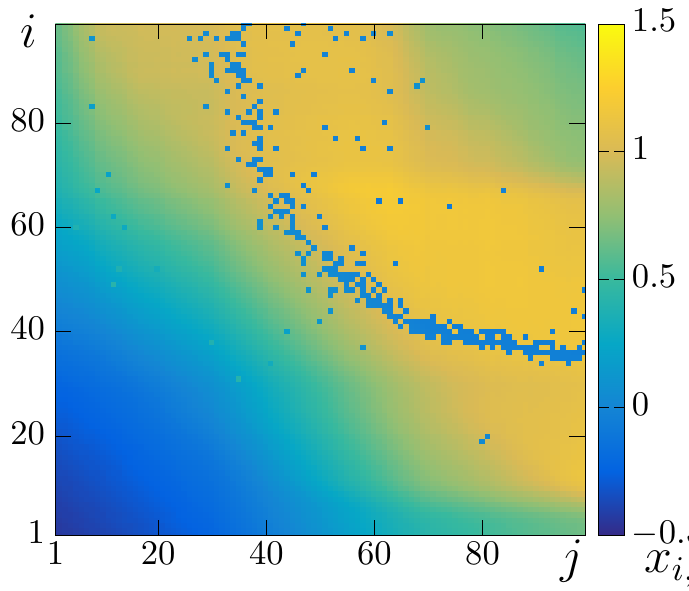}
    \vspace{-9.5mm} \center (a)
\includegraphics[width=\linewidth]{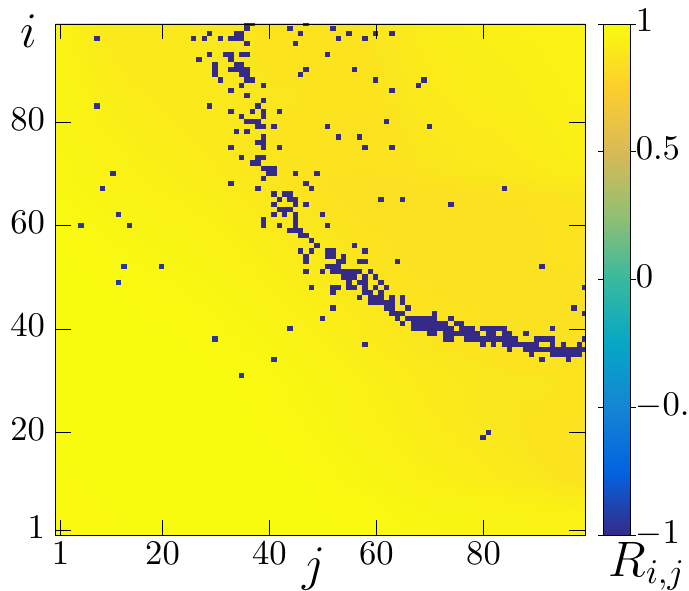}
\vspace{-9.5mm}\center (c)
}
\parbox[c]{.4\linewidth}{
  \includegraphics[width=\linewidth]{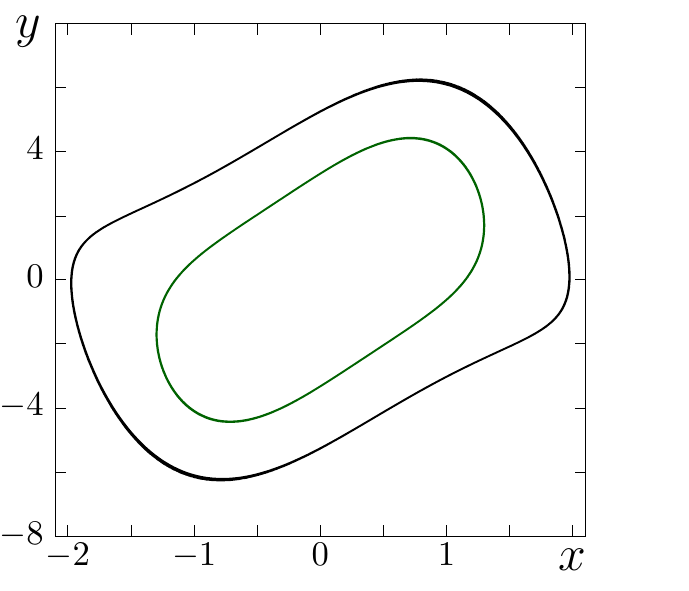}
   \vspace{-9.5mm} \center (b)
\includegraphics[width=\linewidth]{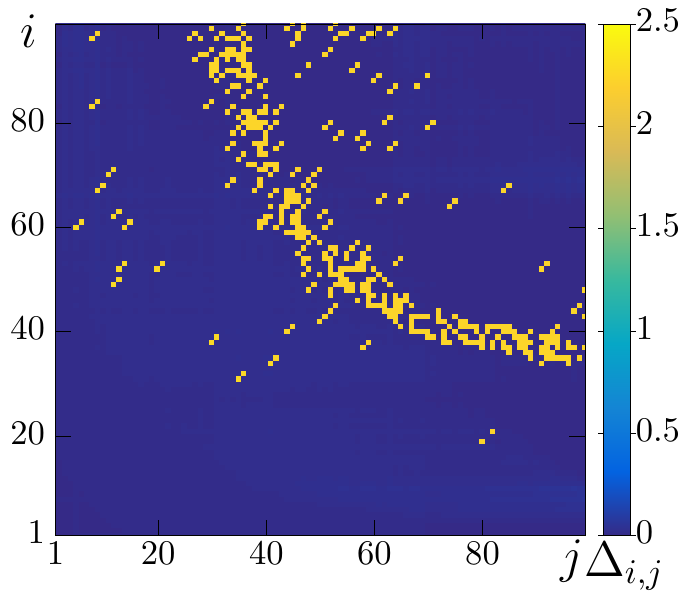}
 \vspace{-9.5mm} \center (d)
}
\caption{(Color online) Solitary state chimera regime (SSC) (region $SSC $ in  the regime diagram in fig.\ref{fig:regime_map}) in the lattice \eqref{eq:grid} for $\sigma=0.6 $ and $r=0.3$.(a) is a snapshot of the system state, (b) phase portrait projections for oscillators with indexes $i=50,~j=45 $ (synchronous region, black line) and  $i=60,~j=45 $(solitary state, green line), (c) presents a spatial distribution of the crosscorrelation $R^{1,1}_{i,j} $ of the oscillator from the synchronous cluster with the other (d) shows spatial distributions of the RMSD $\Delta_{i,j} $. Parameters: $\varepsilon=2 $, $\omega=2.5$, $N=100 $}
\label{fig:SSC}
\end{figure}

An example of this state for another set of randomly distributed initial conditions is shown in fig.\ref{fig:SSC}(a). It is observed that the solitary states form the incoherence cluster, which consists of a large number of solitary states. The region with the synchronous behavior forms the coherence cluster. A few number of solitary states remain randomly distributed within the lattice. 
The regime of solitary state chimera has been found in several systems \cite{rybalova2018mechanism}. However, a type of the chimera state (as well as the solitary state regime) is revealed at first in the models of coupled van der Pol oscillators. The features of the SSC regime are similar to the regime of solitary states. Fig.\ref{fig:SSC}(b) demonstrates the phase portrait projections for elements from the coherence cluster and from the incoherence cluster. It is seen that oscillations from different clusters are characterized by various limit cycles and have the periodic behavior, which is confirmed by a zero value of the maximal Lyapunov exponent. 

The spatial distribution of the crosscorrelation $R^{1,1}_{i,j}$ of an oscillator of the coherence cluster with the others is presented in fig.\ref{fig:SSC}(c). A spatial distribution of the RMSD is shown in fig.\ref{fig:SSC}(d). Oscillators of the coherence cluster are characterized by values of $R^{1,1}_{i,j}$ close to 1 and values of the RMSD close to zero.
 At the same time,  oscillators of the incoherence cluster are characterized by values of the crosscorrelations close to $R^{1,1}_{i,j} \approx -1$ and by the maximal values of the RMSD. 

\begin{figure}[!ht]
\centering
\parbox[c]{.32\linewidth}{ 
  \includegraphics[width=\linewidth]{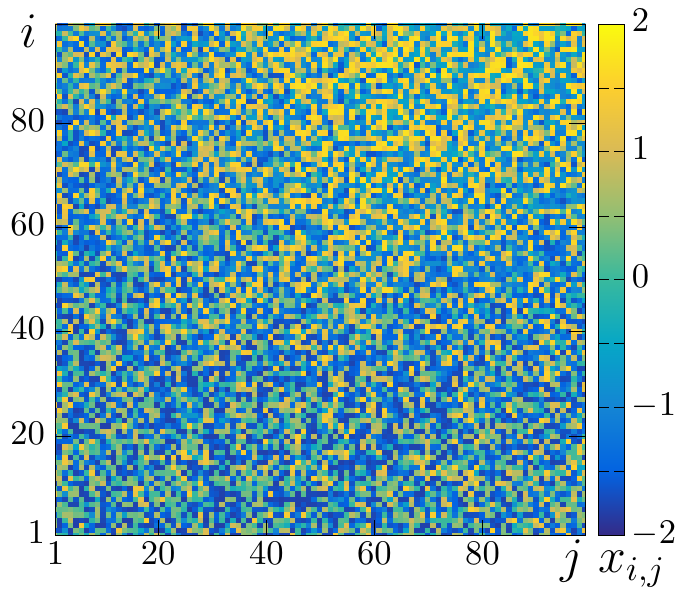}
    \vspace{-9.5mm} \center (a)
}
\parbox[c]{.32\linewidth}{
  \includegraphics[width=\linewidth]{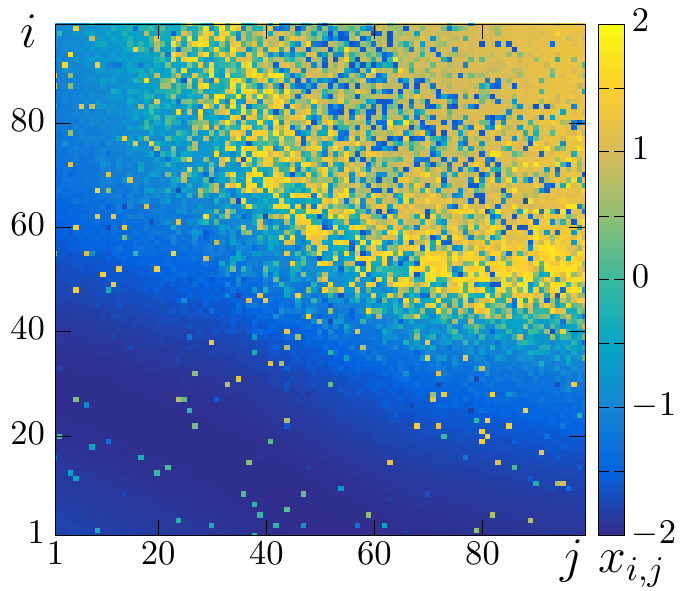}
   \vspace{-9.5mm} \center (b)
}
\parbox[c]{.32\linewidth}{
  \includegraphics[width=\linewidth]{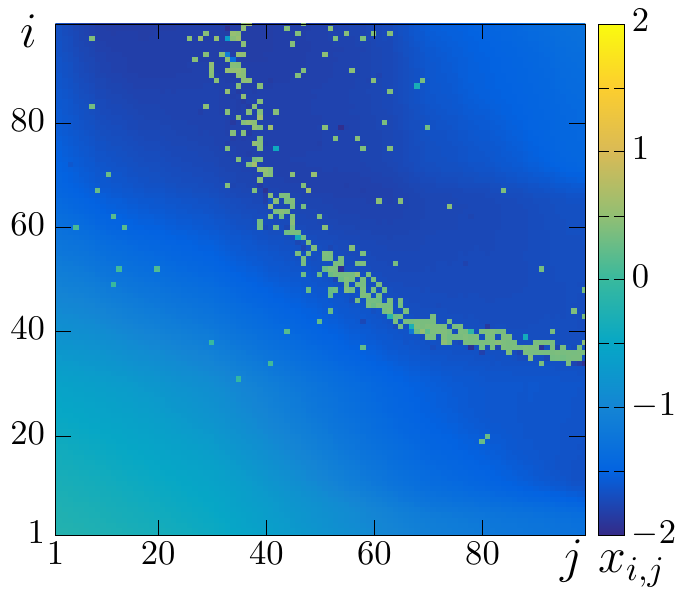}
   \vspace{-9.5mm} \center (c)
}
\caption{(Color online) Establishing the solitary state chimera from the randomly distributed initial conditions. Snapshots of the system state at (a) $t=20 $ time units, (b) $t=40 $, and (c) $t=100 $
Parameters: $\sigma=0.57 $, $r=0.3$ $\varepsilon=2 $, $\omega=2.5$, $N=100 $} 
\label{fig:SSC_evolution}
\end{figure}

We explore the birth of the solitary state chimera from the randomly distributed initial conditions. The evolution of this regime with time is shown in fig.\ref{fig:SSC_evolution}. 
Adjacent elements begin to synchronize with each other after a short enough time (see fig.\ref{fig:SSC_evolution}(a)).  Fig.\ref{fig:SS_evolution}(b) shows that a large part of the oscillators are already synchronized after several time units. 
Incoherence cluster consisting of solitary states is realised at this time. Fig.\ref{fig:SS_evolution}(c) corresponds to the  steady regime, when most of the oscillators are synchronized with each other and incoherence cluster is completely formed and does not change it's spatial location in time. Oscillations of coherence and incoherence clusters are characterized by the corresponding attractors as shown in fig.\ref{fig:SSC}(b). Thus, an establishing process of the solitary state chimera has the same character as that of the solitary state. 

One of the fundamental reason for the appearance of solitary state chimera as well as the solitary state regime is the phenomenon of multistability in the lattice. A start from the randomly distributed initial conditions leads to the fact that a few number of elements are in the basin of attraction of the attractor, which corresponds to the solitary states. Probably the basin of attraction is significantly more narrow than that of the basin of attraction for an attractor in the synchronous region. For this reason most of the oscillators are syncronized during small enough time, while oscillators in the solitary state regime remain  asynchronized. Apparently, increase in the coupling strength leads to significant narrowing of the basin of attraction of the solitary state attractor in a certain spatial region of the lattice. Consequently, solitary states disappears in this region, where the coherence cluster forms.

\section*{Conclusions}

In this work we have investigated a two dimensional lattice network of nonlocally coupled van der Pol oscillators with zero flux boundary conditions. The control parameters of an individual van der Pol oscillator corresponds to the relaxation oscillatory regime. The dynamics of the 2D lattice of vdP oscillators is revealed for both local and nonlocal types of the interaction between oscillators and for relatively low values of the coupling strength ($\sigma \leq 1 $) with fixed control parameters of the oscillators. The nonlocality of coupling adds complexity to the spatiotemporal behavior of the lattice and gives birth to a large number of dynamical regimes. When the nonlocality level is low (small values of the coupling range), the system demonstrates complex spatiotemporal structures, namely spiral wave and target wave chimeras as in \cite{bukh2019bspiral}. 

The features of the spiral wave chimeras are similar to the ones in  the other models \cite{bukh2019bspiral,kuramoto2003rotating,tang2014novel,li2016spiral,shima2004rotating,kundu2018diffusion}. For example, there is no  characteristic bell-like spatial distribution of the mean oscillation frequency. This hints us that a special type of the target wave chimera can be realized in the lattice \eqref{eq:grid}, namely the target wave chimera with a large number of solitary states, which are randomly distributed within the coherence cluster. It is shown that oscillators in the solitary state regime are characterized by an attractor, which differs from the attractor characterized by coherence cluster oscillators.
Moreover, the high-level nonlocality of coupling leads to the appearance of absolutely new spatiotemporal states. The solitary states and solitary state chimera are randomly distributed in space. 

We have constructed the diagram of regimes realized in the lattice \eqref{eq:grid} in the ($r,~\sigma $) parameter plane. It is discovered that the solitary state regime and solitary state chimera exist for the case of long coupling range within a certain interval of the coupling strength, while the target wave chimera is realized for the  same interval of coupling strength and for shorter coupling range. Apparently, all the three dynamical regimes have the similar nature and smoothly transform to each other for different values of the coupling parameters. Increase in the couple range $r $ leads to extension of the wavelength of a target wave chimera (TWC) and for large $r $, the wave process disappears and only solitary states remains inside the synchronous region. On the other hand, growth of the coupling strength leads to grouping of the solitary states in a certain spatial region of the lattice, while the other part becomes completely synchronous without solitary states. Thus, there are two coexisting clusters, namely the coherence cluster with the synchronous behavior of oscillators and the incoherence cluster consisting of a large number of solitary states. This structure have already been found for other systems in \cite{rybalova2018mechanism,rybalova2019solitary}. The solitary states occur due to the bistability (or multistability), which is induced due to the non-local coupling. It is a common mechanism of occurence of chimera state in different systems (see \cite{rybalova2018mechanism}). We present the mechanisms of transition from one regime to the other and offer explanation of disappearance of the solitary states with growth of the coupling strength. 

Besides this, chimera-like structures have been found for the case of very weak coupling strength $\sigma<0.2 $ and short coupling range $r $. This regime is an intermediate state between the complete incoherence, which is observed for the weak local coupling and the synchronous state, in which the system \eqref{eq:grid} for the long coupling range $r $ even for very weak coupling. 

\section*{Acknowledgements}
\vspace{0.5cm}
This work was funded by the Deutsche Forschungsgemeinschaft (DFG, German Research Foundation) -- Projektnummer
163436311-SFB 910. S.S.M. acknowledges the use of New Zealand eScience Infrastructure (NeSI) high performance computing facilities as part of this research.

\bibliographystyle{plain} 
\bibliography{mybib}
\end{document}